\documentclass[prl,twocolumn,10pt,aps]{revtex4-2}
\usepackage[utf8]{inputenc}
\usepackage[normalem]{ulem}
\usepackage{textcomp}
\usepackage{amssymb}
\usepackage{natbib}
\usepackage{amsmath}
\usepackage{amsfonts}
\usepackage{graphicx}
\usepackage{subfigure}
\usepackage{mathrsfs}
\usepackage{dcolumn} 
\usepackage[T1]{fontenc}
\usepackage{bm}
\usepackage{color}
\usepackage{lipsum}
\usepackage{cancel}
\usepackage{mathbbol}
\usepackage{epsfig}
\usepackage{units}
\usepackage{esint}
\usepackage{soul} 
\usepackage{afterpage}
\usepackage{hyperref}
\usepackage{braket}

\usepackage{xurl}
\usepackage{hyperref}
\begin{document}

\title{Vortex screening and the fate of the BKT transition with long-range couplings}

\author{Guido Giachetti}
\email{guido.giachetti@phys.ens.psl.eu}
\affiliation{Laboratoire de Physique de l'\'Ecole Normale Sup\'erieure, CNRS, ENS $\&$ PSL University, Sorbonne Universit\'e, Universit\'e Paris Cité, 75005 Paris, France}

\begin{abstract}
\noindent
The stability of the Berezinskii--Kosterlitz--Thouless phase against long-range interactions remains an open problem. We address this question in the two-dimensional XY model with couplings decaying as $r^{-2-\sigma}$, by developing a real-space renormalization-group treatment near the Gaussian manifold of quasi-long-range ordered (QLRO) states. We retain the full many-body vortex interaction and show that, at leading order in the fugacity, 
screening by small vortex--antivortex pairs preserves its Gaussian-average representation without an expansion in the long-range coupling. Including spin waves perturbatively then yields a local stability criterion: a stable QLRO interval exists only for $\sigma>7/4$. We also investigate a Wilson--Fisher-like fixed point at finite coupling for $\sigma<7/4$, for which the retained flow predicts $\nu^{-1}\sim\sqrt{7/4-\sigma}$ as it merges with the Gaussian line.
\end{abstract}

\maketitle

\paragraph{Introduction.}--- In two dimensions, short-range systems with $O(2)$ symmetry cannot sustain a nonzero order parameter at any finite temperature\,\cite{mermin1966absence, zinnjustin}. They can nevertheless exhibit quasi-long-range order (QLRO), characterized by algebraically decaying correlations\,\cite{berezinskii1971destruction,berezinsky}. The Berezinskii--Kosterlitz--Thouless (BKT) transition provides a natural mechanism for the destruction of QLRO: it can be interpreted as the unbinding of vortex--antivortex pairs and the proliferation of free topological defects\,\cite{KT,K,kosterlitz2017nobel}, which have proven ubiquitous in many fields of physics\,\,\cite{Bishop1978,Rendeira_NatPhys07,bilbro_temporal_2011, Nitsche2014,dalibard_2006,Murthy_PRL115_2015}. Whether this mechanism and the associated QLRO phase survive nonlocal couplings, however, remains actively debated.

Long-range (LR) interactions, decaying as $r^{-d-\sigma}$ in $d$ spatial dimensions, have attracted renewed interest thanks to their increasing experimental accessibility in atomic, molecular, and optical platforms\,\cite{defenu2023long,Monroe2021,haffner2008quantum, saffman2010quantum,browaeys2020many, lahaye2009physics,Yan2013,ritsch2013cold,mivehvar2021cavity}, which provide access to a rich variety of unconventional physical phenomena\,\cite{De2025Nonequilibrium,Zunkovic2018Dynamical,defenu2024out,kastner2011diverging,defenu2021metastability,regemortel2016information,tran2020hierarchy,pappalardi2018scrambling,giachetti2025conditions}. In particular LR interactions can reshape the critical behavior of many-body systems, e.g. by enabling spontaneous symmetry breaking in low dimensions\,\cite{joyce1966spherical,dyson1969existence,thouless1969long,anderson1969exact,cardy1981one}, or by leading to $\sigma$-dependent critical behavior below a crossover threshold $\sigma^*$\,\cite{fisher1972critical,sak1973recursion}. The associated universality classes and the location of this threshold in $O(n)$ models have been investigated
numerically\,\cite{luijten1995monte,luijten1997classical,luijten2001criticality,luijten2002boundary} and through perturbative \,\cite{kosterlitz1976phase, angelini2014relations}, conformal\,\cite{behan2017long,benedetti2025one}, and functional\,\cite{defenu2015fixed,defenu2017criticality,pagni2026one} renormalization group (RG) methods. The crossover to short-range criticality is conventionally described by Sak's criterion\,\cite{sak1973recursion,sak1977low}, $\sigma^*=2-\eta_{\rm sr}$, where $\eta_{\rm sr}$ is the anomalous dimension of the short-range order-parameter correlator. 

\begin{figure*}
\centering
\includegraphics[width=0.8
\linewidth]{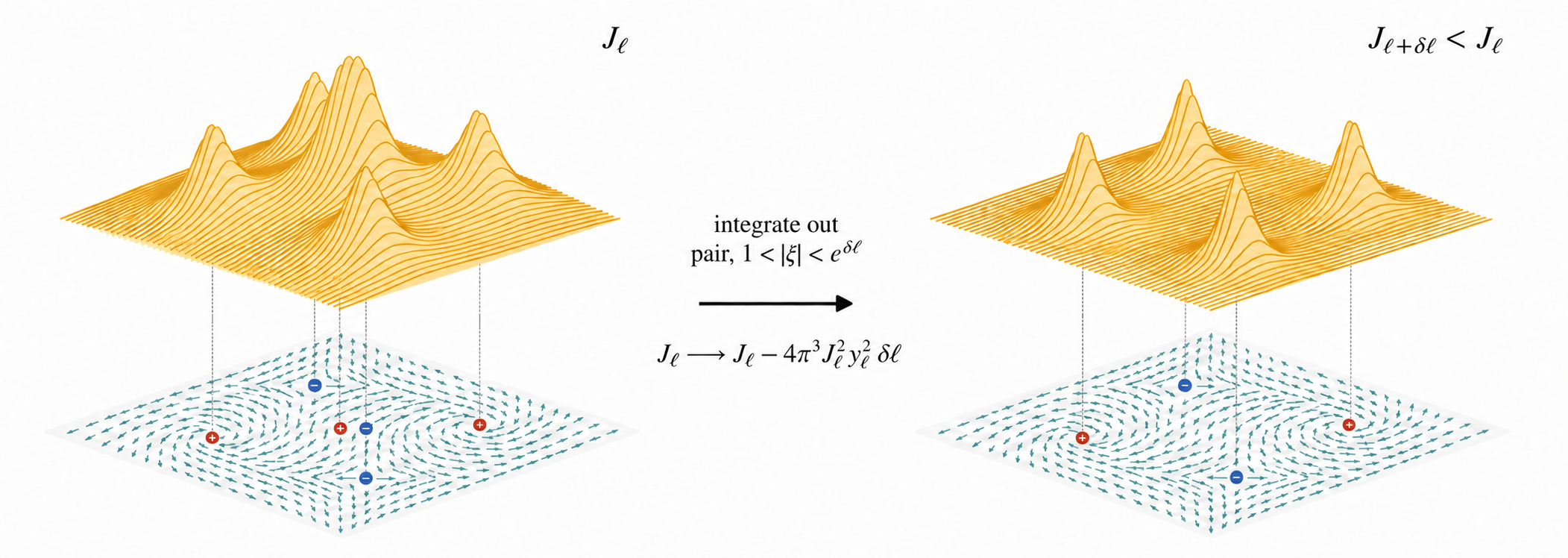} 
\caption{Schematic representation of the renormalization procedure. Spin configurations before (left) and after (right) integrating out a vortex--antivortex pair, with positive and negative vortices marked in red and blue. The yellow surface symbolically represents the many-body vortex interaction. Integrating out the pair screens this interaction by decreasing the spin stiffness $J_\ell$.}
\label{fig:Fig_1}
\end{figure*}
The application of Sak's criterion to the two-dimensional XY model is however particularly subtle because its short-range limit exhibits a QLRO phase with temperature-dependent exponents. In this case thus the location and the very nature of the crossover between the short-range universality and the long-range one remain an open and actively debated problem, attracting renewed attention as recent studies reveal a rich and non-trivial phenomenology.

By means of a perturbative renormalization of the spin-wave contributions, Refs.\,\cite{Giachetti2021,giachetti2022prb} predicted an intermediate QLRO phase between magnetized and disordered phases for $7/4<\sigma<2$\, with related scenarios explored in $U(1)$-symmetric quantum spin chains\,\cite{maghrebi2017continuous}.  On the other hand, Monte Carlo studies of the long-range XY model on a square lattice seem to report direct transitions from the disordered phase to a magnetized phase for $\sigma<2$\,\cite{xiao2025two,yao2025nonclassical}, and have proposed logarithmic QLRO at $\sigma=2$\,\cite{pan2026logarithmic}. Simulations on diluted LR graphs have also yielded contrasting results: early studies reported BKT-like behavior in part of this range, while acknowledging the difficulty of excluding a small spontaneous magnetization\,\cite{ibanez2013critical}; subsequent simulations at $\sigma=1.875$ instead found spontaneous magnetization and no extended scale-invariant phase\,\cite{cescatti2019analysis}. Strong finite-size effects and the potential narrowness of an intermediate QLRO phase complicate the comparison between these scenarios.

Subsequent attempts to include the interplay between vortices and LR couplings have further complicated the picture. The LR generalization of the periodic quadratic Villain
model\,\cite{villain1974quantum,Villain1975theory,jose1977renormalization}, introduced in Ref.\,\cite{giachetti2023villain}, exhibits BKT vortex unbinding for $\sigma >1$ -- a scenario later confirmed in Ref.\,\cite{radzihovsky2024coulomb} -- but differs in universality from the XY scenario. Reference\,\cite{walther2026persistence} proposed instead a deformation of the RG flow of the XY model, which accounts perturbatively for the presence of vortex pairs. The result suggests that an intermediate QLRO phase may persist for arbitrarily small positive $\sigma$\,\cite{walther2026persistence}. The conclusions of Refs.~\cite{giachetti2023villain,walther2026persistence} come from a truncation either of the microscopic vortex Hamiltonian, or of RG-generated couplings, thus relying on potentially uncontrolled approximations.

Our central result is that the screening by small vortex--antivortex pairs preserves the form of the many-body vortex interaction. This allows us to derive a local stability criterion for QLRO, non-perturbative in the form of the vortex interaction, which rules out a stable Gaussian QLRO for every $\sigma<7/4$. The resulting perturbative flow also reveals a finite-coupling critical point below $\sigma=7/4$, suggesting a direct order--disorder transition as the stable QLRO interval disappears.

\paragraph{The model.}--- We consider a two-dimensional lattice of  planar rotors, interacting through the ferromagnetic $O(2)$-symmetric Hamiltonian
\begin{equation} \label{eq:model}
\beta H =  \frac{1}{2} \sum_{\mathbf{i} \neq \mathbf{j}} J(r) \left[ 1 -  \cos(\theta_{\mathbf{i}}-\theta_\mathbf{j}) \right] \, , 
\end{equation}
where $r = |\mathbf{i} - \mathbf{j} | $ while $\mathbf{i} ,\mathbf{j} \in \mathbb{Z}^2$ and  $\theta_\mathbf{i}$ represent the lattice rotor variables. We are here implicitly measuring the distance in units of the lattice spacing. We assume for the coupling $J(r) \sim g\, r^{-2-\sigma}$, for $r \gg 1$, with $\sigma > 0$ in order to ensure additivity of the thermodynamic potentials\,\cite{campa2014physics}. 

For low enough temperatures, we can restrict ourselves to configurations that are smooth (up to a finite number of isolated singularities) for which $\theta_\mathbf{j}$ can be replaced by a continuous function $\theta(\mathbf{x})$. While for $r \sim 1$ a gradient expansion $1 - \cos (\theta_{\mathbf{i}}-\theta_\mathbf{j}) \sim |\nabla \theta|^2$ is justified, 
the same approximation is not justified, even on smooth configurations, for the power-law tail which couples far-away points. We thus write $\beta H = S_{\rm sr} [\theta] + S_{g} [\theta]$ with 
\begin{equation} \label{eq:SsrSg}
\begin{split}
    S_{\rm sr}[\theta] &=  \frac{J}{2} \int d^2\mathbf{x} \ | \nabla \theta|^2 , \\
    S_{\rm g} [\theta] &=  \frac{g}{2} \int_{r >1} \frac{d^2\mathbf{x} \ d^2\mathbf{x}^\prime}{r^{2+\sigma}} \left[ 1 -  \cos(\theta(\mathbf{x})-\theta(\mathbf{x}^\prime)) \right]  ,\, 
\end{split}
\end{equation}
where $J>0$ is an effective coupling which accounts for the short-range part of $J(r)$. 
This is the action that appears in Refs.\,\cite{Giachetti2021,giachetti2022prb,Giachetti2021EPL,walther2026persistence}. 

\paragraph{Integration of the vortex pairs.}--- As for the usual BKT transition, the field $\theta(\mathbf{x})$ appearing in Eq.\,\eqref{eq:SsrSg} is not a single-valued function as, for each point, it is defined up to multiples of $2 \pi$, and thus topological configurations can become relevant. Such configurations have a nonzero integer winding of $\theta(\mathbf{x})$ around one or more points on the plane. It can be shown (see End Matter) that any possible configuration of $\theta(\mathbf{x})$ can be decomposed as 
\begin{equation} \label{eq:decomposition}
    \theta(\mathbf{x}) = \theta_{\rm sw} (\mathbf{x})  + \sum_j m_j \arg(\mathbf{x} -\mathbf{x_j}) \, , 
\end{equation}
where $\theta_{\rm sw}(\mathbf{x})$ is a generic single-valued function, $\arg(\mathbf{x})$ is the (multi-valued) argument function on the plane and $m_j \in \mathbb{Z}$ can be interpreted as point-like charges each located in $\mathbf{x}_j$. The decomposition \eqref{eq:decomposition} has a clear physical meaning, namely the decomposition into a spin-wave contribution $\theta_{\rm sw} (\mathbf{x})$ and a vortex one, due to point-like integer topological charges. In the following we will retain only elementary charges $m_j = \pm1$, corresponding to vortices and antivortices, since higher charges are suppressed at low temperature. 

As in the standard BKT construction\,\cite{KT}, in the short-range action $S_{\rm sr}$, these two contributions factorize. For a globally neutral configuration of $p$ vortices and $p$ antivortices, $S_{\rm sr}^{(p)} [\theta] = S_{\rm sw}[\theta_{\rm sw}] + S^{(p)}_{C}[m]$ where
\begin{equation} \label{eq:Ssr}
\begin{split}
    S_{\rm sw} [\theta_{\rm sw}]&= \frac{J}{2}\int d^2\mathbf{x}\, |\nabla \theta_{\rm sw}|^2 , \\
    S^{(p)}_{C} [m] &= - \pi J \sum_{i \neq j}^{2p} m_i m_j \ln |\mathbf{x}_i - \mathbf{x}_j| ,
\end{split}
\end{equation}
while the non-neutral configurations (i.e. those for which $\sum_j m_j \neq 0$) have an infinite energy cost, and are thus forbidden. 

The factorization between spin waves and topological excitations is however no longer valid in the long-range Hamiltonian, since the long-range cosine is not quadratic. For a configuration of $p$ vortex--antivortex pairs, we therefore define the core-subtracted long-range action as $S^{(p)}_g[\theta]=S_g[\theta]-2p\, \varepsilon_c$, with $\theta$ given by Eq.\,\eqref{eq:decomposition}. The core energy $\varepsilon_{c}$ has been introduced so that the zero of $S^{(p)}_{g}$ corresponds to the configuration with $p$ mutually distant vortex-antivortex pairs, each of separation $|\boldsymbol{\xi}|=1$. Let us notice that for these configurations $S^{(p)}_{C} = 0$ as well.

Thus, the vortex gas partition function for a long-range XY model takes the form 
\begin{equation}  \label{eq:Zsp}
    Z =  \sum_{p=0}^\infty \frac{y^{2p}}{p!^2} \int \prod_{j=1}^{2p} d^2\mathbf{x}_j \int \mathcal{D}[\theta_{\rm sw}] \, \, e^{-S^{(p)}_{\rm sr}[\theta]} \,  e^{- S_{g}^{(p)}[\theta]}  \, ,
\end{equation}
where $y$ is the fugacity of the vortex gas, and $p!^2$ avoids the overcounting of the pairs. The main difference between $S_g$ and its short-range counterpart is that, because of Eq.\,\eqref{eq:decomposition}, the former contains $n$-body interactions among vortices with arbitrary $n$. 
Indeed, as shown in the Supplementary Material, truncating the long-range Hamiltonian \eqref{eq:SsrSg} while retaining the topological contribution gives back the long-range Villain model studied in Ref.\,\cite{giachetti2023villain}, which is known to belong to a different universality class.

\paragraph{Vortex-renormalization.}--- We follow here the established procedure of Ref.\,\cite{KT}, namely we gradually integrate out vortex-antivortex pairs with separation $1 < |\boldsymbol{\xi}|< e^{\delta \ell}$ and reabsorb their contribution as a screening to the action (see Fig.\,\ref{fig:Fig_1} for a sketch of the procedure). 

While the details of the renormalization procedure are left to the Supplementary Material, we will discuss here the main result, namely that, in spite of its apparent intricacy, the form of Eq.\,\eqref{eq:Zsp} is preserved by this renormalization scheme. Remarkably, indeed, the screening contribution by small vortex--antivortex pairs preserves the Gaussian-average representation of Eq.\,\eqref{eq:Zsp} at leading order in the fugacity and within the dipole approximation, regardless of the exact form of $S_g$. In particular, as in the short-range Coulomb gas\,\cite{KT,kosterlitz2017nobel}, the effect of the integration of small vortex pairs results in an effective screening of the stiffness $J$. The physical intuition behind this result is that small vortex pairs generate logarithmic screening analogous to that of the Coulomb potential, which can be interpreted as an additional Gaussian average coming from a free-massless action of the same form as $S_{\rm sr}$ in Eq.\,\eqref{eq:SsrSg}, thereby renormalizing $J$. 

As in the standard BKT treatment, we get $\delta J=-4\pi^3J^2y^2\delta\ell$, while the corresponding change in the fugacity, due to the redefinition of the zero of the $p$-pairs action, is given by
\begin{equation} \label{eq:deltay}
    \delta y = y\delta\ell\left[-\pi J+\frac12\left.\partial_\xi\ln\left\langle e^{-S_g[\theta_{\rm sw}+\theta_\xi]}\right\rangle_{\rm sw}\right|_{\xi=1}\right].
\end{equation}
Here $\theta_\xi$ represents the field of a vortex-pair at a distance $\xi$, while $\langle \cdot \rangle_{\rm sw}$ denotes the average over the Gaussian action $S_{\rm sw}[\theta_{\rm sw}]$ in Eq.\,\eqref{eq:Ssr}. 

While spin-wave integration instead generates additional operators and thus does not preserve the form of Eq.\,\eqref{eq:SsrSg}, the vortex-screening result nevertheless applies also to the renormalized interaction, since its derivation does not assume a specific form of $S_g$. In this representation thus no explicit $O(y^2)$ contribution to the flow of the couplings in $S_g$ can appear, as the screening is absorbed into the running of $J$ and $y$.

\paragraph{Stability of the QLRO phase.}--- To discuss the stability of the Gaussian QLRO phase ($g=y=0$) we have to include the contribution of the fast spin waves to complete the RG step. In this regime $g \ll J$ we can perform this integration perturbatively, following Ref.\,\cite{Giachetti2021}. To first order in $g$, $\langle e^{-S_g} \rangle_{\rm sw}\simeq e^{-\langle S_g \rangle_{\rm sw}}$, so that the spin-wave correlations preserve the form of the action, up to a dressing in the coupling $r^{-\eta_J} = \langle \cos (\theta_{\rm sw}(\mathbf{x})-\theta_{\rm sw}(\mathbf{x'})) \rangle_{\rm sw}$, with $\eta_J=1/(2\pi J_\ell)$. The scaling dimension of $g$ becomes thus $2-\Delta_J$, where $\Delta_J=\sigma+\eta_J$. In particular, the correction to the fugacity can be expressed in terms of the spin-wave-averaged energy $\left\langle S_g[\theta_{\rm sw} +\theta_\xi]\right\rangle_{\rm sw}$ of a single vortex--antivortex pair. This quantity is estimated in the Supplementary Material
\begin{equation} \label{eq:Landau-Peierls}
    \left\langle S_g[\theta_{\rm sw} +\theta_\xi]\right\rangle_{\rm sw} = \frac{g\pi^2 \xi^2}{\Delta_J^2} \left( 1- \Delta_J \ln \frac{\xi}{4} \right) \, .
\end{equation}
The resulting flow is given by
\begin{equation}  \label{eq:RG_4eqs}
    \begin{split}
    \dot{J}_\ell &= - 4 \pi^3 y_\ell^2 J_\ell^2 + \frac{\pi}{2} g_\ell \, \, \\
    \dot{y}_\ell &= \left[2 -  \pi J_\ell - \frac{\pi^2 g_\ell}{2\Delta_{J_\ell}^2} (2  + (4\ln 2-1)\Delta_{J_\ell}) \right] y_\ell  , \\
     \dot{g}_\ell &= (2 - \Delta_{J_\ell}) g_\ell  \\
    \end{split}
\end{equation}
Here the contribution $(\pi/2)g$ to the stiffness flow comes from the change of the cutoff in $S_{g}$. 

\begin{figure*} 
\centering
\includegraphics[width=
\linewidth]{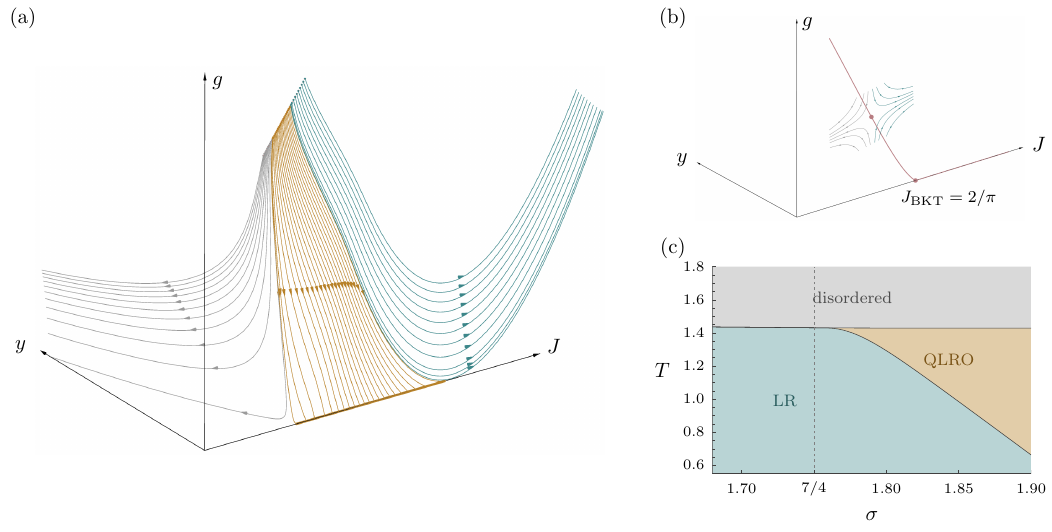} 
\caption{RG flows and phase structure. (a) Trajectories at $\sigma=1.85$, colored according to their LR (teal), QLRO (ochre), or disordered (gray) basin. (b) Fixed-point locus as $\sigma$ varies (pink): the LR-marginal endpoint of the Gaussian line, $J=[2\pi(2-\sigma)]^{-1}$, $g=y=0$, for $7/4<\sigma<2$, joins the finite-coupling branch for $\sigma<7/4$ at $J_{\rm BKT}=2/\pi$. The nearby saddle flow is schematic. (c) Illustrative numerical RG phase diagram for $J_0=1/T$, $g_0=10^{-4}/T$, and $y_0=e^{-6/T}$, with the same color code.}
\label{fig:Fig_2}
\end{figure*}

The RG flow \eqref{eq:RG_4eqs} establishes a competition between long-range ordering $g_\ell \to \infty$ (low-temperature phase) and vortex proliferation $y_\ell \to \infty$ (high-temperature phase), while the possible intermediate QLRO phase corresponds to $y=g=0$  (see Fig.\,\ref{fig:Fig_2}(a) for a sketch of the flow). As in Ref.~\cite{Giachetti2021}, this Gaussian line is stable against both perturbations when $2-\sigma<\eta_{\bar J}<1/4$ (see Fig.\,\ref{fig:Fig_2}), a condition that cannot be met for $\sigma=7/4$. While always accessible for sufficiently small $g$ and $y$, the intermediate QLRO phase can become very narrow on a set of physical initial conditions (see Fig.\,\ref{fig:Fig_2}(c)), while finite-coupling corrections may cause microscopic trajectories to bypass its basin of attraction. 

Although Eq.\,\eqref{eq:RG_4eqs} remains perturbative in $g$ and $y^2$, it retains the vortex interaction at every many-body order. This differs from the truncated flow in  Ref.~\cite{walther2026persistence} (see Supplementary Material for a comparison). A nontrivial independent check of the leading $O(g)$ corrections to the flow comes from a conformal calculation at $\Delta_J=2$, which in particular reproduces the coefficient $\pi^2\ln2$ in $\dot{y}/y$ obtained above (see End Matter).

\paragraph{Finite-coupling fixed point.}--- For $\sigma<7/4$, Eq.\,\eqref{eq:RG_4eqs} also admits a fixed point at finite $g$ and $y$, given by $J_*=1/(2\pi(2-\sigma))$, $g_*\propto (7-4\sigma)$, $y_*^2\propto (7-4\sigma)$, which merges with the extreme of the Gaussian line at $J_*=2/\pi$ as $\sigma \to 7/4$ (see Fig.\,\ref{fig:Fig_2}(b)). The linearization of the retained flow gives one relevant and two irrelevant directions, supporting its interpretation as a Wilson--Fisher-like critical point separating the ordered and disordered phases. Its positive eigenvalue yields
\begin{equation}
\nu^{-1} \sim \sqrt{7/4-\sigma}.
\end{equation}
While additional second-order terms, which become non-negligible for $\sigma \sim 7/4$, are needed to determine the numerical prefactors consistently, the square-root scaling is expected to persist. This scaling parallels that of the one-dimensional long-range Ising model as its interaction approaches $1/r^2$\,\cite{kosterlitz1976phase,cardy1981one,benedetti2025one}.

\paragraph{Conclusions.}--- In this work we obtain a stability criterion for the Gaussian QLRO manifold in the long-range XY model, non-perturbative in the form of the many-body vortex action, by retaining the full many-body interaction, and reabsorbing vortex screening exactly, at leading order in $y^2$, in the renormalization of the spin stiffness. Close to QLRO this allows us to determine the local stability conditions, which place the lower threshold for an open interval of stable Gaussian fixed points at $\sigma=7/4$. 

Our analysis differs substantially from previous treatments of topological charges in long-range $O(2)$-symmetric models~\cite{giachetti2023villain,walther2026persistence}, which suggest QLRO persistence at smaller value of $\sigma$, as the full form of the vortex interaction is preserved under renormalization to the retained order. This provides a self-consistent description of vortex screening and its effect on local stability, without requiring additional generated couplings to be set to zero. 

The finite-coupling fixed point found for $\sigma<7/4$ in the retained flow suggests a description of the direct order--disorder transition near the onset of QLRO.  The contrasting numerical findings on diluted LR graphs and on the square lattice\,\cite{ibanez2013critical,cescatti2019analysis,xiao2025two,yao2025nonclassical} may reflect the narrowness of the intermediate phase, or the effect of additional strong-coupling fixed points for $\sigma>7/4$, which could limit the QLRO basin without changing its local stability. Our formulation offers a route to investigate this possibility through a systematic treatment of spin-wave contributions to $S_g$ and the flow at finite $g$. Near $\sigma=7/4$, conformal perturbation theory around the Gaussian compact-boson theory could provide a systematic starting point for this analysis. Numerical studies targeting the vicinity of the QLRO manifold, aided by recent advances in long-range Monte Carlo algorithms\,\cite{mueller2026efficient}, would provide a complementary test. 

\paragraph{Acknowledgments.}--- The author acknowledges the support of the European MSCA Grant 101152898 (DREAMS).

\section{End Matter}

\subsection{Proof of Eq.\,\eqref{eq:decomposition}}
As any configuration on the plane must have an integer, but in general non-zero, winding number, $\nabla \theta$ is not a conservative field, but rather, for a generic closed path $\gamma$ in the plane, $\oint_\gamma \nabla \theta \cdot d \boldsymbol{\ell} = 2 \pi M$, with $M \in \mathbb{Z}$. As $M$ is quantized, the only way to satisfy this condition is $\nabla \times \nabla \theta = 2 \pi \sum_{j} m_j \delta (\mathbf{x}-\mathbf{x}_j)$ with $m_j \in \mathbb{Z}$. By the Helmholtz theorem, we can thus write 
\begin{equation} \label{eq:gradient_decomposition}
    \nabla \theta = (\partial_x \theta_{\rm sw} - \partial_y \psi, \partial_y \theta_{\rm sw} + \partial_x \psi) \, ,
\end{equation}
where $\theta_{\rm sw}(\mathbf{x})$ and $\psi(\mathbf{x})$ are single-valued functions, with 
\begin{equation}
    \nabla^2 \psi = 2 \pi \sum_{j} m_j \delta (\mathbf{x}-\mathbf{x}_j)    
\end{equation}
We thus get $\psi(\mathbf{x}) = \sum_j m_j \ln |\mathbf{x}-\mathbf{x}_j|$. By introducing the complex coordinate $z = x+iy$, we get $\psi(z) = \text{Re}\sum_j m_j \ln (z-z_j)$ and, by Cauchy-Riemann, 
\begin{equation}
\begin{split}
    (- \partial_y \psi, \partial_x \psi) &= \nabla \sum_j m_j \text{Im} \ln (z- z_j) \\ &= \sum_j m_j \nabla \arg(z-z_j) \, . 
\end{split}
\end{equation}
By replacing this in Eq.\,\eqref{eq:gradient_decomposition} we get our result.

\subsection*{Fugacity flow at $\Delta_J = 2$}
We show how to recover the $O(g)$ correction in the flow at $\Delta_J=2$, through conformal arguments. Let us place a vortex--antivortex pair at the complex coordinates $0$ and $\xi$: the $O(g)$ energy cost is
\begin{equation}
\frac g2\int_{|z-z'|>1}\frac{d^2z\,d^2z'}{|z-z'|^4}\left[1-\cos(\arg\chi)\right],
\end{equation}
where $\chi=z(z'-\xi)/((z-\xi)z')$. The conformal map $w=z/(\xi-z)$, followed by $v=w'/w$, gives $\chi=1/v$ and transforms the measure into $d^2w\,d^2v/(|w|^2|1-v|^4)$. Near $v=1$, the original cutoff becomes $|1-v|>\epsilon(w)$, with $\epsilon(w)=|1+w|^2/(\xi|w|)$. The integral over $v$ can be directly evaluated
\begin{equation}
\int_{|1-v|>\epsilon}\frac{d^2v}{|1-v|^4}\left[1-\cos(\arg v)\right]
\sim \frac{\pi}{2}\ln\frac1\epsilon+\pi\ln2.
\end{equation}

On the other hand, the angular average of $\ln[1/\epsilon(w)]$ is $\ln\xi-|\ln|w||$. The logarithmically growing contributions therefore give, with $t= \ln |w|$,
\begin{equation}
\begin{aligned}
\pi g\int_{-\ln\xi}^{\ln\xi}dt\left[\frac{\pi}{2}(\ln\xi-|t|)+\pi\ln2\right]\\
=\frac{\pi^2g}{2}(\ln\xi)^2+2\pi^2g\ln2\,\ln\xi .
\end{aligned}
\end{equation}
Posing $\ln\xi = \delta \ell$, the term $\pi^2g/2 (\ln\xi)^2$ can be encoded in the stiffness flow $\dot J=\pi g/2$. The remaining term yields $c=\pi^2\ln2$, and hence
\begin{equation}
\frac{\dot y}{y}=2-\pi J-\pi^2g\ln2.
\end{equation}

\clearpage
\newpage
\newpage
\setcounter{equation}{0}
\setcounter{page}{1}
\renewcommand{\theequation}{S.\arabic{equation}}
\setcounter{figure}{0}
\renewcommand{\thefigure}{S.\arabic{figure}}

\begin{widetext}

\begin{center}
\textbf{\Large Supplemental Material: \\
\, \\
Vortex screening and the fate of the BKT transition with long-range couplings} \\ 
\vspace{0.5cm}
Guido Giachetti \\
\emph{Laboratoire de Physique de l'\'Ecole Normale Sup\'erieure, CNRS, ENS $\&$ PSL University, Sorbonne Universit\'e, Universit\'e Paris Cité, 75005 Paris, France}
\end{center}

\section{Villain Regime}
\noindent
In this section we show that the quadratic truncation of the cosine action $S_g$ in \eqref{eq:SsrSg} of the main text, namely
\begin{equation}
    S_g \simeq \frac{g}{4} \int \frac{d^2 \mathbf{x} \ d^2\mathbf{x}^\prime}{|\mathbf{x}-\mathbf{x}^\prime|^{2 +\sigma}} \left( \theta(\mathbf{x}) - \theta(\mathbf{x}^\prime)\right)^2
\end{equation}
is the continuum analogue of the long-range Villain model introduced in \cite{giachetti2023villain} on the lattice, as it gives back the same vortex-vortex interaction once the spin waves have been integrated out.

Let us exploit the decomposition \eqref{eq:decomposition} of the main text, namely
\begin{equation}
    \theta(\mathbf{x}) = \theta_{\rm sw}(\mathbf{x}) + \sum_j m_j \arg(\mathbf{x}- \mathbf{x}_j) \, ,
\end{equation}
and let us also assume that the quadratic short-range spin-wave term $S_{\rm sw}[\theta]$ is not relevant (i.e. we set $J=0$). For a $p$-vortex-pair configuration we have 
\begin{equation} 
\begin{split}
     S_{g}^{(p)}[\theta_{\rm sw},m] \simeq 
    &\frac{g}{4} \int \frac{d^2 \mathbf{x} \ d^2\mathbf{x}^\prime}{|\mathbf{x}-\mathbf{x}^\prime|^{2 +\sigma}} \left( \theta_{\rm sw} (\mathbf{x})- \theta_{\rm sw} (\mathbf{x}^\prime)  - \sum_{j=1}^{2p} m_j  \Psi_{\mathbf{x}_j}(\mathbf{x},\mathbf{x}^\prime) \right)^2 \, 
\end{split}
\end{equation}
where $\Psi_{\mathbf{x}_j} (\mathbf{x},\mathbf{x}^\prime) = \arg(\mathbf{x'}- \mathbf{x}_j) - \arg(\mathbf{x}- \mathbf{x}_j)$ corresponds to the oriented angle corresponding to the segment $\mathbf{x'}-\mathbf{x}$, seen by the point $\mathbf{x}_j$. Following the procedure of \cite{giachetti2023villain}, the spin-wave modes can be integrated out exactly. Introducing the Fourier
kernel
\begin{equation}
\mathcal{K} (q) = \int d^2\mathbf{r}\,r^{-2-\sigma}\left[1-\cos(\mathbf{q}\cdot\mathbf{r})\right] 
\end{equation}
we obtain the effective action 
\begin{equation}
\begin{split}
S_{g}^{(p)} [m]  = \int \frac{d^2 \mathbf{r} \ d^2\mathbf{r}^\prime}{(r^\prime r)^{2 +\sigma}}  \int d^2 \mathbf{x} \ d^2\mathbf{x}^\prime f(|\mathbf{x}-\mathbf{x}'|)
\left( \sum_j m_j\,\mathcal{C}_j(\mathbf{x},\mathbf{r},\mathbf{r}') \right)
\left(\sum_{j'}m_{j'}\,\mathcal{C}_{j'}(\mathbf{x}',\mathbf{r},\mathbf{r}')\right),
\end{split}
\end{equation}
where
\begin{equation}
f(x) = \frac{g}{16} \int\frac{d^2\mathbf{q}}{(2\pi)^2}\, \frac{e^{i\mathbf{q}\cdot\mathbf{x}}}{\mathcal{K} (q)}.
\end{equation}
and $\mathcal{C}_j(\mathbf{x},\mathbf{r},\mathbf{r}')$ denotes the circulation of $\Psi_{\mathbf{x}_j}$ around the parallelogram
$\mathcal{P}(\mathbf{x};\mathbf{r},\mathbf{r}')$ of vertices $\mathbf{x}$, $\mathbf{x}+\mathbf{r}$, $\mathbf{x}+\mathbf{r}+\mathbf{r}^\prime$, $\mathbf{x}+\mathbf{r}^\prime$, namely
\begin{equation}
\begin{split}
\mathcal{C}_j(\mathbf{x},\mathbf{r},\mathbf{r}')=
\Psi_{\mathbf{x}_j}(\mathbf{x},\mathbf{x}+\mathbf{r}) + \Psi_{\mathbf{x}_j}(\mathbf{x}+\mathbf{r},\mathbf{x}+\mathbf{r}+\mathbf{r}') +
\Psi_{\mathbf{x}_j}(\mathbf{x}+\mathbf{r}+\mathbf{r}',\mathbf{x}+\mathbf{r}') + \Psi_{\mathbf{x}_j}(\mathbf{x}+\mathbf{r}',\mathbf{x}) .
\end{split}
\end{equation}
Recalling that $\Psi_{\mathbf{x}_j}(\mathbf{x},\mathbf{x}')$ corresponds to the oriented angle corresponding to the segment $\mathbf{x'}-\mathbf{x}$, seen by the point $\mathbf{x}_j$, such circulation is 
\begin{equation}
\mathcal{C}_j(\mathbf{x},\mathbf{r},\mathbf{r}') =2\pi\,\operatorname{sgn}(\mathbf{r}\times\mathbf{r}') 
\end{equation}
if $\mathbf{x}_j$ lies inside the parallelogram, while $\mathcal{C}_j= 0$ otherwise (we are assuming that $\mathbf{x}_j$ does not lie on the boundary of the parallelogram). Thus
\begin{equation} \label{eq:VillainSm}
\begin{split}
S_{g}^{(p)} =& 4\pi^2  \int \frac{d^2 \mathbf{r} \ d^2\mathbf{r}^\prime}{(r^\prime r)^{2 +\sigma}}  \int d^2 \mathbf{x} \ d^2\mathbf{x}^\prime f(|\mathbf{x}-\mathbf{x}'|)
\sum_{\substack{\mathbf{x}_j\in
\mathcal{P}(\mathbf{x};\mathbf{r},\mathbf{r}')}} \sum_{
\mathbf{x}_{j'}\in
\mathcal{P}(\mathbf{x}';\mathbf{r},\mathbf{r}')}
m_j m_{j'}\, ,
\end{split}
\end{equation}
which can be rewritten as a generalized Coulomb gas Hamiltonian
\begin{equation}
    S_{g}^{(p)} = -\sum_{i \neq j}^{2 p} m_i m_{j} U(|\mathbf{x}_i - \mathbf{x}_j|) 
\end{equation}
where we introduce the vortex-vortex potential
\begin{equation}
    U(|\mathbf{x}_i-\mathbf{x}_j|) = 4\pi^2  \int \frac{d^2 \mathbf{r} \ d^2\mathbf{r}^\prime}{(r^\prime r)^{2 +\sigma}}  \int_{\mathcal{A}(\mathbf{x},\mathbf{x}')}  d^2 \mathbf{x} d^2 \mathbf{x}^\prime f(|\mathbf{x}-\mathbf{x}'|).
\end{equation}
where $\mathcal{A}(\mathbf{x},\mathbf{x}') = \lbrace \mathbf{x},\mathbf{x'} | \mathbf{x}_i \in \mathcal{P}(\mathbf{x},\mathbf{r},\mathbf{r'}) \mathbf{x}_j \in \mathcal{P}(\mathbf{x'},\mathbf{r},\mathbf{r'}) \rbrace$. 

The long-distance behavior of such potential has been examined in Ref.\,\cite{giachetti2023villain}. In particular, for $1 < \sigma <2$, fixing $U(1)=0$, we have the power law behavior
\begin{equation}
    U(r) \sim \frac{g}{\sigma-1} (r^{2-\sigma} - 1) 
\end{equation}
while for $\sigma < 1$ the effective coupling of the potential diverges, signaling that the topological configurations cannot be excited. Here we have exploited the global neutrality of the gas, to choose the zero of the potential in $r=1$. 

While in the infrared, the bare vortex--antivortex interaction grows as a power law with separation, seemingly preventing vortex unbinding at any finite temperature, Ref.\,\cite{giachetti2023villain} shows that the screening by smaller bound pairs modifies its long-distance form, producing an effective logarithmic interaction in the infrared.
Vortices can therefore unbind through a BKT mechanism $\forall \sigma \in (1,2)$, in spite of the stronger confinement suggested by the bare potential. This does not imply full equivalence with the short-range XY model: the low-temperature phase has finite magnetization, rather than merely quasi-long-range order.

\section{Vortex pair screening}
\noindent
We will now explicitly carry out the real space renormalization group on the partition function $Z$ in Eq.\,\eqref{eq:Zsp} of the main text, namely, 
\begin{equation}
\label{eq:Zp_incomplete}
Z=
\sum_{p=0}^{\infty}
\frac{y^{2p}}{p!^2}
\int\prod_{j=1}^{2p}d^2\mathbf{x}_j
\int\mathcal{D}[\theta_{\rm sw}]\,
e^{-S^{(p)}[\theta]} \, , 
\end{equation}
where $\theta(\mathbf{x})$ is a generic configuration with $p$ charges $m_j=+1$ and $p$ charges $m_j=-1$, and can be decomposed into spin waves and topological excitations, according to Eq.\,\eqref{eq:decomposition} of the main text:
\begin{equation}
\theta(\mathbf{x})= \theta_{\rm sw}(\mathbf{x})+ \sum_{j=1}^{2p}m_j\arg(\mathbf{x}-\mathbf{x}_j) \, .
\end{equation} 
Here $S[\theta]=S^{(p)}_{\rm sr}[\theta]+S^{(p)}_g[\theta]$, where 
\begin{align}
   S_{\rm sr}[\theta]&= \frac{J}{2}\int d^2\mathbf{x}\, |\nabla\theta(\mathbf{x})|^2, \, 
\end{align}
or, equivalently,  
\begin{equation} \label{eq:Gaussian+Coulomb}
S_{\rm sr}[\theta]=\frac J2\int d^2\mathbf{x}\,|\nabla\theta_{\rm sw}(\mathbf{x})|^2-\pi J\sum_{i\ne j}m_i m_j\ln|\mathbf{x}_i-\mathbf{x}_j|,
\end{equation}
In this section, we do not assume any particular form for $S_g[\theta]$. In the UV the form of $S_g[\theta]$ is anyway given in the main text
\begin{equation}
    S^{(p)}_g[\theta] = \frac{g}{2} \int_{r>1}
\frac{d^2\mathbf{x}\,d^2\mathbf{x}'}{r^{2+\sigma}}\left[1-\cos\bigl(\theta(\mathbf{x})-\theta(\mathbf{x}')\bigr)\right] - 2 p \,\varepsilon_c \, .
\end{equation}

\subsection*{Integration of the close pairs}
\noindent
Here we integrate out vortex--antivortex pairs with separation $1<|\boldsymbol{\xi}|<e^{\delta\ell}$. We will show that, for small $y$ (and for a dilute vortex gas) this simply renormalizes the spin stiffness $J$ as 
\begin{equation}
    J \to J - 4 \pi^3 y^2 J^2 \, \delta \ell
\end{equation}
regardless of the form of $S_g[\theta]$ and the value of the couplings.
\\\\
In the limit $\delta\ell\to0$, we can integrate out one pair at a time.  Let us denote the positive and negative charges of the vortex pair we integrate by $\mathbf{x}_\pm$ and parameterize 
\begin{equation}
\mathbf{x}_{\pm}= \mathbf{R}\pm\frac{\boldsymbol{\xi}}{2}.
\end{equation}
For a dilute gas, the interaction with distant charges can be expanded in the pair separation. We retain the leading dipolar contribution, quadratic in $\boldsymbol{\xi}$. 

The contribution to the field $\theta(\mathbf{x})$ coming from the pair we want to integrate is given by
\begin{align}
\delta\theta(\mathbf{x}) &= \arg(\mathbf{x}-\mathbf{R} -\boldsymbol{\xi}/2) -\arg(\mathbf{x}-\mathbf{R}+\boldsymbol{\xi}/2)
\\
&=\boldsymbol{\xi}\cdot (\mathbf{\hat{z}} \times 
\nabla_{\mathbf R}\ln|\mathbf{x}-\mathbf R|)
+O(\xi^3) \equiv \boldsymbol{\xi}\cdot \mathbf{u} (\mathbf{x},\mathbf{R}) +O(\xi^3) .
\label{eq:dipole_phase_shift}
\end{align}
We have thus $S^{(p)} = S^{(p-1)}+ \Delta S + 2 \varepsilon_c$, where $\Delta S = \xi_\alpha a_\alpha+\xi_\alpha \xi_\beta b_{\alpha \beta}/2+O(\xi^3)$ with 
\begin{equation}
    \begin{split}
        a_\alpha &= \int d^2\mathbf{x}\,\frac{\delta S_g}{\delta\theta(\mathbf{x})}u_\alpha(\mathbf{x},\mathbf R)-2\pi J\sum_{j=1}^{2p-2}m_j\nabla_{R_\alpha}\ln|\mathbf{x}_j-\mathbf R|,\\
b_{\alpha \beta}  &=  \int d^2\mathbf{x}\,d^2\mathbf{x}'\,\frac{\delta^2S_g} {\delta\theta(\mathbf{x})\delta\theta(\mathbf{x}')} u_\alpha (\mathbf{x},\mathbf R) u_\beta (\mathbf{x}^\prime,\mathbf R).
    \end{split}
\end{equation}
Now the $p$-vortex-pair term ($p\geq1$) of the sum in Eq.\,\eqref{eq:Zp_incomplete} can be rewritten as 
\begin{align}
\frac{y^{2p}p^2}{p!^2} \int\prod_{j=1}^{2p-2}d^2\mathbf{x}_j
\int\mathcal D[\theta_{\rm sw}]\, e^{-S^{(p-1)}}  \int_{1<|\boldsymbol{\xi}|<e^{\delta\ell}}
d^2\mathbf R\,d^2\boldsymbol{\xi}\, e^{-\Delta S} , 
\end{align}
where the combinatorial factor $p^2$ comes from the possible choices of the pair to integrate among the $p$ positive and $p$ negative charges, and the term $ \epsilon_c$ has been reabsorbed in the common normalization of $Z$. In turn, this can be rearranged as
\begin{align}
\frac{y^{2(p-1)}}{(p-1)!^2} \int\prod_{j=1}^{2p-2}d^2\mathbf{x}_j
\int\mathcal D[\theta_{\rm sw}]\, e^{-S^{(p-1)}} y^2\delta\ell \int d^2\mathbf R\int_{\xi =1} d \xi \, e^{-\Delta S}
\end{align}
At leading order in $y^2$, this renormalizes the remaining-sector action as
\begin{equation}
\delta S^{(p-1)} = -y^2\delta\ell \int d^2\mathbf R \int_{\xi =1} d \xi \, e^{-\Delta S}.
\end{equation}
Expanding $\Delta S$ in the dipolar distance and performing the angular integration we have 
\begin{equation}
\int_{\xi =1} d \xi \, e^{-\Delta S} =
2\pi +\frac{\pi}{2} \bigl(|\mathbf a|^2-\operatorname{tr}b\bigr)
+O(\xi^4).
\end{equation}
and, dropping the field-independent contributions, 
\begin{equation}
\label{eq:dipole_action_variation}
\delta S^{(p-1)} =
\frac{\pi y^2\delta\ell}{2}
\int d^2\mathbf R\, \left[\operatorname{tr}b-|\mathbf a|^2\right].
\end{equation}
The Coulomb and long-range terms in $\mathbf a$ give independent contribution to the flow after the integration over $\mathbf{R}$, as the cross term contains 
\begin{equation}
    -\int d^2 \mathbf{R}\,  \mathbf{u} \cdot \nabla_\mathbf{R} \sum_j m_j \ln|\mathbf{x}_j -\mathbf{R}| = \int d^2 \mathbf{R}\,  \nabla_\mathbf{R} \ln|\mathbf{x} -\mathbf{R}| \times \nabla_\mathbf{R} \sum_j m_j \ln|\mathbf{x}_j -\mathbf{R}|
\end{equation}
which is a total derivative and vanishes at the boundary. We thus obtain
\begin{align} 
\delta S= 4\pi^4J^2 y^2\delta\ell \sum_{i\ne j} m_i m_j \ln |\mathbf{x}_i - \mathbf{x}_j| +\frac12\int d^2\mathbf{x}\,d^2\mathbf{x}'\,\delta C(\mathbf{x},\mathbf{x}')\left[\frac{\delta^2S_g}{\delta\theta(\mathbf{x})\delta\theta(\mathbf{x}')}-\frac{\delta S_g}{\delta\theta(\mathbf{x})}\frac{\delta S_g}{\delta\theta(\mathbf{x}')}\right]\, ,
\label{eq:deltaS_functional}
\end{align}
where
\begin{equation}
    \begin{split}
        \delta C(\mathbf{x},\mathbf{x}') &= \pi y^2\delta\ell
\int d^2\mathbf R\, \, \mathbf{u}(\mathbf{x}) \cdot \mathbf{u}(\mathbf{x}') =\pi y^2\delta\ell \int d^2\mathbf R\, \nabla_{\mathbf R} \ln|\mathbf{x}-\mathbf{R}| \cdot  \nabla_{\mathbf R} \ln|\mathbf{x'}-\mathbf{R}| \\&= 2 \pi^2 y^2 \delta\ell (\ln L - \ln |\mathbf{x}-\mathbf{x'}|)\, .
    \end{split}
\end{equation}
Here $L$ is the IR cutoff of the system. The terms $\propto \ln L$ drop out by $O(2)$ invariance under a uniform translation of $\theta$. After dropping the constant contribution proportional to $\ln L$, we can write
\begin{equation}
\delta C(\mathbf{x},\mathbf{x}')=  4\pi^3J^2y^2\delta\ell \, \,\partial_JG(\mathbf{x},\mathbf{x}'),
\end{equation}
where $G(\mathbf{x},\mathbf{x}')=\eta_{J}\ln|\mathbf{x}-\mathbf{x}'| = 1/2 \langle (\theta(\mathbf{x}) - \theta (\mathbf{x}'))^2\rangle $ is the two-point function computed on the Gaussian spin-wave theory \eqref{eq:Gaussian+Coulomb}. We therefore obtain
\begin{equation} \label{eq:deltaS_final}
\delta S=\delta J \left[\pi\sum_{i\ne j}m_i m_j\ln|\mathbf{x}_i-\mathbf{x}_j|-\frac12 e^{S_g} \int d^2\mathbf{x}\,d^2\mathbf{x}'\,\partial_JG(\mathbf{x},\mathbf{x}')\frac{\delta^2 e^{-S_g}}{\delta\theta(\mathbf{x})\delta\theta(\mathbf{x}')} \right].
\end{equation}
with $\delta J = 4\pi^3J^2y^2\delta\ell$. 

Both contributions correspond to the same change of $J \to J - \delta J$ in $S_{\rm sr}$: the first straightforwardly renormalizes the Coulomb-gas part of the action; the second instead renormalizes the spin waves. This can be seen explicitly by taking the Boltzmann factor of the second term in Eq.\,\eqref{eq:deltaS_final}
\begin{equation}
\begin{split}
     \int \mathcal{D}[\theta_{\rm sw}] \, e^{-S_{\rm sw} - S_g - \delta S} \to &\int \mathcal{D}[\theta_{\rm sw}] \, e^{-S_{\rm sw}}\left( 1 + \frac{\delta J}{2} \int d^2\mathbf{x}\,d^2\mathbf{x}'\,\partial_J G(\mathbf{x},\mathbf{x}') \frac{\delta^2 }{\delta\theta(\mathbf{x})\delta\theta(\mathbf{x}')} \right) e^{-S_g} \\
     &\propto \int \mathcal{D}[\theta_{\rm sw}] \, e^{-S_{\rm sw}} e^{+\delta J/2 \int |\nabla \theta_{\rm sw}|^2} e^{-S_g} = \int \mathcal{D}[\theta_{\rm sw}] \, e^{-(J-\delta J)/2 \int |\nabla \theta_{\rm sw}|^2} e^{-S_g} \, .
\end{split}
\end{equation}

\subsection{Fugacity renormalization}
\noindent
We now examine the contribution to the RG flow coming from the spin-wave integration, to first order in $g$. For $p$ mutually distant vortex--antivortex pairs of separation $\xi$, neglecting their residual interactions, the reference free-energy cost is $p$ times that of a single pair. We denote the field of one pair by $\theta_\xi(\mathbf{x})=\arg(\mathbf{x}-\xi\hat{\mathbf e}_x/2)-\arg(\mathbf{x}+\xi\hat{\mathbf e}_x/2)$. Up to $\xi$-independent normalization factors, the corresponding reference fugacity satisfies
\begin{equation}
y^{2p}\propto\left\langle e^{-S[\theta_{\rm sw}+\theta_\xi]}\right\rangle_{\rm sw}^{p},
\end{equation}
where $\langle\cdot\rangle_{\rm sw}$ denotes the normalized average over the Gaussian spin-wave action $S_{\rm sr}[\theta_{\rm sw}]$. This contribution is subtracted from the action in the exponential because it is already included in the Gaussian measure. Thus,
\begin{equation}
\ln y=\frac12\ln\left\langle e^{-S[\theta_{\rm sw}+\theta_\xi]}\right\rangle_{\rm sw}.
\end{equation}
Changing the reference separation from $\xi$ to $\xi e^{\delta\ell}$ therefore gives
\begin{equation}
\frac{\delta y}{y}=\frac{\delta\ell}{2}\left.\partial_\xi\ln\left\langle e^{-S[\theta_{\rm sw}+\theta_\xi]}\right\rangle_{\rm sw}\right|_{\xi=1},
\end{equation}
with the derivative taken at fixed couplings and cutoffs. Separating the short-range contribution, $S_{\rm sr}[\theta_{\rm sw}+\theta_\xi]-S_{\rm sr}[\theta_{\rm sw}]=2\pi J\ln\xi$, up to $\xi$-independent terms, we obtain
\begin{equation}
\frac{\delta y}{y}=\delta\ell\left[-\pi J+\frac12\left.\partial_\xi\ln\left\langle e^{-S_g[\theta_{\rm sw}+\theta_\xi]}\right\rangle_{\rm sw}\right|_{\xi=1}\right].
\end{equation}

\section{Single-pair energy cost}
\noindent
We now compute the $O(g)$ expression of the energy cost \eqref{eq:Landau-Peierls} of the main text, namely
\begin{equation}
\left\langle S_g[\theta_{\rm sw}+\theta_\xi]\right\rangle_{\rm sw}
=\frac g2\int_{r>1}\frac{d^2\mathbf{x}\,d^2\mathbf{x}'}{r^{2+\sigma+\eta_{J}}}\left[1-\cos(\theta_\xi(\mathbf{x})-\theta_\xi(\mathbf{x}'))\right]\,.
\end{equation}
In the following $\Delta_J = \eta_J + \sigma$ as in the main text. Since $\theta_\xi (\mathbf{x})$ is the field of a neutral pair, it becomes small at large distances, while it is singular close to the vortex pair. On the other hand, we can write
\begin{align} \label{eq:decomposizione_magica}
1-\cos(\theta_\xi(\mathbf{x})-\theta_\xi(\mathbf{x}'))
={}&(1-\cos\theta_\xi(\mathbf{x}))+(1-\cos\theta_\xi(\mathbf{x}')) - 4 \cos \frac{\theta_\xi(\mathbf{x})- \theta_\xi(\mathbf{x'})}{2} \sin \frac{\theta_\xi(\mathbf{x})}{2} \sin \frac{\theta_\xi(\mathbf{x}')}{2} \, .
\end{align}
Let us examine the third term. Its contribution to the integral is negligible in the regions close to the vortex cores: indeed, if e.g. $\mathbf{x}$ is close to one of the charges, $\mathbf{\theta}(\mathbf{x}')$ is almost constant, while $\theta(\mathbf{x})$ winds around the core and viceversa. Thus, in the integral we can approximate 
\begin{equation}
    4 \cos \frac{\theta_\xi(\mathbf{x})- \theta_\xi(\mathbf{x'})}{2} \sin \frac{\theta_\xi(\mathbf{x})}{2} \sin \frac{\theta_\xi(\mathbf{x}')}{2} \simeq (\boldsymbol{\xi} \cdot \nabla \theta (\mathbf{x})) (\boldsymbol{\xi} \cdot \nabla \theta (\mathbf{x'})) =  \frac{\xi^2}{2} \nabla \ln|\mathbf{x}| \cdot \nabla \ln |\mathbf{x'}| \, 
\end{equation}
(as the crossed derivatives vanish in the integral). By setting $\mathbf{x'} = \mathbf{x} + \mathbf{r}$ and exploiting the Green identity   
\begin{align}
\int_{|\mathbf{x}|<R} \nabla \ln|\mathbf{x}| \cdot \nabla \ln |\mathbf{x+r}|d^2\mathbf{x}\,&=2\pi(\ln R-\ln r),
\end{align}
the contribution to $\left\langle S_g[\theta_{\rm sw}+\theta_\xi]\right\rangle_{\rm sw}$ coming from the third term can be written as
\begin{equation}
     \frac{g}{2} \xi^2 \int_{r>1} \frac{d^2\mathbf{r}}{r^{2+\Delta_J}}  \left( \pi \ln r - \pi \ln R \right)
\end{equation}
The integral of the contribution $1-\cos\theta_\xi(\mathbf{x})$ in Eq.\,\eqref{eq:decomposizione_magica} over the same region $|\mathbf{x}| < R$, can instead computed exactly resorting to elliptic coordinates with foci at the two vortices,
\begin{equation}
x_1=\frac{\xi}{2}\cosh\mu\cos\nu,\qquad x_2=\frac{\xi}{2}\sinh\mu\sin\nu.
\end{equation}
The energy cost and the area element are
\begin{equation}
1-\cos\theta_\xi(\mathbf{x})=\frac{2\sin^2\nu}{\sinh^2\mu+\sin^2\nu},
\qquad
d^2\mathbf{x}=\frac{\xi^2}{4}(\sinh^2\mu+\sin^2\nu)\,d\mu\,d\nu.
\end{equation}
At the outer boundary $|\mathbf{x}|=R$, we have $\mu=\ln(4R/\xi)+O(\xi^2/R^2)$. Therefore,
\begin{equation}
\int_{|\mathbf{x}|<R}d^2\mathbf{x}\,[1-\cos\theta_\xi(\mathbf{x})]
=\frac{\xi^2}{2}\int_0^{2\pi}d\nu\,\sin^2\nu\int_0^{\ln(4R/\xi)}d\mu
=\frac{\pi\xi^2}{2}\ln\frac{4R}{\xi} \, . 
\end{equation}
An identical result holds for the second $\mathbf{x} \to \mathbf{x'}$ term, so that their total contributions in $\left\langle S_g[\theta_{\rm sw}+\theta_\xi]\right\rangle_{\rm sw}$ is
\begin{equation}
    \frac{g}{2} \xi^2 \int_{r>1} \frac{d^2\mathbf{r}}{r^{2+\Delta_J}}   \left( \pi \ln (4R) - \pi \ln \xi \right) \, .
\end{equation}
Finally, sending $R \to \infty$ we have 
\begin{equation}
    \left\langle S_g[\theta_{\rm sw}+\theta_\xi]\right\rangle_{\rm sw} = g \pi^2 \xi^2 \int_1^\infty \frac{dr}{r^{1+\Delta_{J}}} \ln \frac{4r}{\xi} = \frac{g \pi^2 \xi^2}{\Delta_J^2}  \left( 1+\Delta_J\ln(4/\xi)\right)
\end{equation}

\section{Relation with other RG descriptions}
\noindent
Here we compare our treatment with the published RG calculation of Ref.~\cite{walther2026persistence}. While both approaches retain the leading contribution from integrating out a vortex pair in a dilute gas, they encode the flow in different couplings. Our calculation derives screening directly from the full many-body vortex interaction, preserving its Gaussian-average representation without truncating the interaction within the retained dipolar approximation. Reference~\cite{walther2026persistence} instead extracts running parameters from the response to a uniform twist. The leading normalization matching is
\begin{equation}
\beta=J,\qquad y_{\rm W}=2y,\qquad g_{\rm W}=\frac{g}{2J}.
\end{equation}
With these identifications, Eq.~(9) of Ref.~\cite{walther2026persistence} reproduces our stiffness flow at orders $g$ and $y^2$. However, Eq.~(C11) of Ref.~\cite{walther2026persistence} also generates an additional coupling $\gamma$, which is subsequently set to zero although $\gamma=0$ is not generally invariant under the flow. Our vortex-pair integration requires no analogous truncation of a generated coupling.

A further distinction concerns the behavior near marginality. The fugacity flow of Ref.~\cite{walther2026persistence} becomes
\begin{equation}
\frac{\dot y_{\rm W}}{y_{\rm W}}=2-\pi J+\frac{\pi^2g}{2(2-\Delta_J)},
\end{equation}
whereas our estimate gives
\begin{equation}
\frac{\dot y}{y}=2-\pi J-c(J,\sigma)g,
\qquad
c(J,\sigma)=\pi^2\left[\frac{1}{\Delta_J^2}+\frac{2\ln2-\tfrac12}{\Delta_J}\right].
\end{equation}
Both corrections suppress vortex proliferation for $J>2/\pi$, but the coefficient in Ref.~\cite{walther2026persistence} diverges at the marginal point $\Delta_J \to 2$, while ours approaches the finite value $\pi^2\ln2$. This value is further confirmed by the conformal calculation in the End Matter.

This pole reflects a nonuniform separation into effective parameters, rather than a divergent pair energy at fixed separation: Eq.~(B5) of Ref.~\cite{walther2026persistence} contains the regular combination $(\xi^{2-\Delta_J}-1)/(2-\Delta_J)\to\ln\xi$. Singular coefficients also occur at the BKT boundary in Eq.~(9) of Ref.~\cite{walther2026persistence}. Our retained flow remains regular at both QLRO boundaries, allowing their local stability to be assessed without these nonuniform parameter identifications.

\section*{References}

\begingroup
\small
\begin{list}{}{%
    \setlength{\leftmargin}{2.8em}%
    \setlength{\labelwidth}{2.3em}%
    \setlength{\labelsep}{0.5em}%
    \setlength{\itemsep}{0.4em}%
    \setlength{\parsep}{0pt}%
    \setlength{\topsep}{0pt}%
}

\item[{[\citenum{giachetti2023villain}]}]
G.~Giachetti, N.~Defenu, S.~Ruffo, and A.~Trombettoni,
Villain model with long-range couplings,
\href{https://doi.org/10.1007/JHEP02(2023)238}
{J. High Energy Phys. \textbf{2023} (2), 238}.

\item[{[\citenum{walther2026persistence}]}]
L.~Walther, J.~Willsher, and J.~Knolle,
Persistence of the Berezinskii-Kosterlitz-Thouless Transition
with Long-Range Couplings,
\href{https://doi.org/10.1103/9y2v-ybdb}
{Phys. Rev. Lett. \textbf{136}, 227102 (2026)}.

\end{list}
\endgroup

\end{widetext}
\end{document}